\documentclass[12pt]{article}
\usepackage{epsfig,amsfonts,amssymb}
\usepackage{hyperref}
\usepackage{cite}
\input epsf.sty
\usepackage{cite}

\def\ZZZ{{\hbox{ Z\kern-1.6mm Z}}}
\def\RRR{{\hbox{ R\kern-2.4mm R}}}
\def\CCC{{\hbox{ C\kern-2.0mm C}}}
\def\zzz{{\hbox{z\kern-1mm z}}}

\newcommand{\qeq}{{\hbox{=\kern-2.3mm ? \kern.5mm }}}
\renewcommand{\qeq}{=}

\newcommand{\vp}{\varphi}

\newcommand{\GG}{{\cal G}}

\newcommand{\HH}{{\cal H}}

\newcommand{\wt}{\widetilde}
\newcommand{\wh}{\widehat}

\newcommand{\be}{\begin{equation}}
\newcommand{\ee}{\end{equation}}
\newcommand{\ben}{\begin{eqnarray}\displaystyle}
\newcommand{\een}{\end{eqnarray}}

\newcommand{\refb}[1]{(\ref{#1})}
\newcommand{\p}{\partial}
\newcommand{\sectiono}[1]{\section{#1}\setcounter{equation}{0}}

\def\one{{\hbox{ 1\kern-.8mm l}}}
\def\zero{{\hbox{ 0\kern-1.5mm 0}}}

\newcommand{\bea}[1]{\begin{eqnarray}\label{#1} }
\newcommand{\eea}{\end{eqnarray}}

\newcommand{\eqref}{\refb}

\usepackage{bm}
\usepackage[table]{xcolor}

\newcommand{\cL} {\{\hskip -4pt\{}
\newcommand{\cR} {\}\hskip -4pt\}}

\begin{document}

\baselineskip 24pt

\begin{center}
{\Large \bf  Wilsonian Effective Action of Superstring Theory}

\end{center}

\vskip .6cm
\medskip

\vspace*{4.0ex}

\baselineskip=18pt

\centerline{\large \rm Ashoke Sen}

\vspace*{4.0ex}

\centerline{\large \it Harish-Chandra Research Institute}
\centerline{\large \it  Chhatnag Road, Jhusi,
Allahabad 211019, India}

\centerline{and}

\centerline{\large \it Homi Bhabha National Institute}
\centerline{\large \it Training School Complex, Anushakti Nagar,
    Mumbai 400085, India}

\vspace*{1.0ex}
\centerline{\small E-mail:  sen@mri.ernet.in}

\vspace*{5.0ex}

\centerline{\bf Abstract} \bigskip

By integrating out the heavy fields in type II or heterotic string field theory one
can construct the effective  action for the light fields. This effective theory
inherits all the algebraic structures  of the parent theory and the effective action
automatically satisfies the
Batalin-Vilkovisky quantum
master equation. This theory
is manifestly ultraviolet finite, has only light 
fields as its explicit degrees of freedom,
and the Feynman diagrams of this theory reproduce the
exact scattering amplitudes of light states in string theory to any arbitrary order in
perturbation theory. Furthermore in
this theory the degrees of freedom of light fields above certain energy 
scale
are also implicitly 
integrated out. This energy scale is determined by a particular parameter
labelling a family of equivalent actions, 
and can be made arbitrarily low, leading to the interpretation
of the effective action as the Wilsonian effective action.

\vfill \eject

\baselineskip 18pt

\tableofcontents

\sectiono{Introduction} \label{s1}

The original motivation for string field theory was to develop tools for
studying non-perturbative aspects of string theory. 
For open string field theory this goal has been partially realized in the form of
non-trivial classical solutions that are not accessible to perturbation theory\cite{0511286 }.
Similar success for closed strings has not been forthcoming despite some tantalizing
numerical results\cite{0506077}.
Nevertheless superstring field theory can provide us with
practical tools for addressing issues that arise within perturbation theory, {\it e.g.}
in dealing with situations where the masses of external states get renormalized or
the classical vacuum gets destabilized and new stable vacua arise near the original
classical vacuum\cite{1508.02481}. 

Superstring field theory has infinite number of fields with masses going all the way
up to infinity.
We may want to consider a
situation where the only states that participate in the scattering are `light' particles 
-- those that
are massless at string tree level -- but we want to include the effect of higher derivative
corrections as well as loop corrections where heavy string states propagate in the
loop. 
In this case the most useful description will be provided by an effective action
of light fields obtained by integrating out the heavy state contributions. This
effective action will still be capable of reproducing the full string theory amplitudes
involving light external states to any arbitrary order in perturbation theory. In 
particular, like the parent string field theory, this effective field theory will be free
from ultraviolet (UV) divergences. Our
goal in this paper will be to analyze the structure of such an effective action.

Our main result is that the 
interaction vertices of the effective field theory so constructed satisfy the same
algebraic identities as those of the parent string field theory\cite{1508.05387}. It
then follows that the effective action satisfies the Batalin-Vilkovisky (BV) master
equation and can be quantized using the BV 
formalism\cite{bv,thorn,hata1,hata,9206084}. 
Furthermore following the analysis of \cite{br1,br2} 
one can show that
in this effective action the modes of the light fields above a certain energy scale
are also integrated out implicitly. This energy scale is determined 
by a parameter, known as the stub length,
that the effective action inherits from the original
action\cite{sonoda,9205088,9301097}. By varying this parameter we can
change this energy scale, which also determines the effective UV cut-off of the theory.
Therefore the effective action so constructed can be regarded as the Wilsonian
effective action\cite{wilson1,wilson2,polchinski} 
of superstring field theory. 

Given this effective action, one can use it to compute any quantity involving the light
fields, {\it e.g.} S-matrix elements, using standard Feynman rules.  
The absence of UV 
divergences is made manifest by the exponential suppression of the vertices at large
euclidean momenta provided one takes the ends of the 
integration contours over loop energies
to be at $\pm i\infty$\cite{1604.01783}. 
Furthermore as long as the energies
of the external states remain below the threshold of production of the 
heavy particles, the
loop energy integration contours can be chosen to avoid
the singularities in the vertices arising from the  effect of integrating 
out the heavy particles.

It is also possible to construct the one
particle irreducible (1PI) effective action of the light fields by also integrating out the
light fields propagating in the loops. This satisfies a different set of properties
which can also be derived using the same general procedure. Such an effective
action can be useful for finding the correct vacuum, and computing the renormalized
masses of light fields around this vacuum. For example the entire
analysis of \cite{1508.02481} could be carried out using such an effective action
without having to deal with the heavy string states. The final result of course will
remain unchanged.

The rest of the paper is organized as follows. In \S\ref{s2} we review some of the
algebraic structures that arise in the conventional superstring field theory, and derive
some Ward identities for the off-shell amplitudes of this theory. In \S\ref{s3} we
integrate out the heavy fields of the theory and use manipulations similar to
those in \S\ref{s2} to derive certain identities involving the interaction vertices of this
effective field theory. These identities are then used to show that the resulting 
effective action
satisfies the BV master equation. We also describe the construction of the 1PI action
of the light fields and show that this satisfies the classical master equation.
In \S\ref{s4} we describe the role of the stub length
parameter in controlling the energy scale above which the modes of the light
fields in the effective action
are also implicitly integrated out. We conclude in \S\ref{s6} with a discussion 
of our results and possible generalizations where, instead of integrating out all the heavy modes
of the string, we integrate out the modes above a certain mass level and construct an effective
theory of string fields below that mass level. In appendix \ref{sb} we describe explicitly the
algorithm for constructing the vertices of the effective action using off-shell amplitudes of light
states only.
In appendix \ref{s5} we describe how 
a set of spurious fields which are present in the effective 
action do not play any role in 
the evaluation of Feynman diagrams.

\sectiono{Algebraic structures in superstring field theory} \label{s2}

We shall first review some algebraic structures that appear in the 
construction of superstring field theories\cite{1508.05387} generalizing the corresponding 
results in closed bosonic string field theory\cite{sonoda,9206084}.
We shall denote by $\HH_T$
the subspace of GSO even states in the matter ghost superconformal field theory (SCFT)
satisfying
\ben \label{ech}
&& b_0^- |s\rangle =0, \quad L_0^- |s\rangle=0, \quad \hbox{for $|s\rangle\in \HH_T$}
\, ,
\een
where  
\be
b_0^\pm =b_0\pm \bar b_0, \quad c_0^\pm ={1\over 2} (c_0\pm \bar c_0), \quad
L_0^\pm = L_0\pm \bar L_0\, .
\ee
$L_n,\bar L_n$ are the total Virasoro generators, and
$b_n,\bar b_n,c_n, \bar c_n$ are the modes of the usual $b,\bar b,c,\bar c$ 
ghost fields.  We do not make any assumption about the background except that
it is described by some arbitrary world-sheet SCFT.
For heterotic string theory we shall denote by
$\HH_m$ the subspace
of states in $\HH_T$ carrying picture number $m$, whereas for type II string theories
$\HH_{m,n}$ will denote the subspace
of states in $\HH_T$ carrying left-moving picture number $m$ and right-moving picture
number $n$. We also define
\ben 
\hbox{for heterotic} &:& \wh\HH_T \equiv \HH_{-1}\oplus \HH_{-1/2}, \quad
\wt\HH_T\equiv \HH_{-1}\oplus \HH_{-3/2} \, ,
\nonumber \\ 
\hbox{for type II} &:&
\cases{\wh\HH_T \equiv \HH_{-1,-1}\oplus \HH_{-1/2,-1}\oplus \HH_{-1,-1/2}\oplus \HH_{-1/2,-1/2}\cr
\wt \HH_T \equiv \HH_{-1,-1}\oplus \HH_{-3/2,-1}\oplus \HH_{-1,-3/2}\oplus \HH_{-3/2,-3/2}}\, .
\een
For both heterotic and type II string theories we take 
$|\vp_r\rangle\in \wh\HH_T$, $|\vp_r^c\rangle \in
\wt\HH_T$
to be appropriate basis states satisfying
\be \label{ebas}
\langle\vp_r^c |c_0^- | \vp_s\rangle =\delta_{rs}\, ,\quad 
\langle\vp_r |c_0^- | \vp_s^c\rangle =\delta_{rs}\, .
\ee
The second relation follows from the first.
\refb{ebas} implies the completeness relation
\be\label{ecom}
\sum_r |\vp_r\rangle \langle \vp_r^c|c_0^- = {\bf 1}, \quad 
\sum_r |\vp_r^c\rangle \langle \vp_r|c_0^- = {\bf 1}\, ,
\ee
acting on states in $\wh\HH_T$ and $\wt\HH_T$ respectively.
The basis states $\vp_r$ and
$\vp_r^c$ will in general carry non-trivial grassmann parities which we shall denote
by $(-1)^{\gamma_r}$ and $(-1)^{\gamma_r^c}$ respectively. 
In the NS 
sector of the heterotic theory and the NSNS and RR sector of type II
theory, the grassmann parity of $\vp_r$ or $\vp_r^c$ is odd (even) if the ghost number
of $\vp_r$  or $\vp_r^c$ is odd (even). In the R sector of the heterotic theory and the
RNS and NSR sector of the type II theory, the grassmann parities of the states will
have opposite correlation with the ghost number. 
It follows from the
ghost number conservation rule and \refb{ebas} that
\be \label{egrrule}
(-1)^{\gamma_r+\gamma_r^c}=-1\, .
\ee

For heterotic string theories we shall denote by $\GG$ the 
identity operator in the NS sector and the 
zero mode
of the picture changing operator (PCO) in the R sector. 
For type II string theories $\GG$ will be defined as the 
identity operator in the NSNS sector, zero mode of the left-moving picture
number in the RNS sector, zero mode of the right-moving picture
number in the NSR sector, and product of the zero modes of the left-moving
and right-moving picture numbers in the RR sector. 
$\GG$ satisfies
\be \label{egrel}
[\GG, L_0^\pm]=0, \qquad [\GG, b_0^\pm]=0\, , \qquad [\GG, Q_B]=0\, ,
\ee
where $Q_B$ is the BRST charge.

Using the standard identification 
between the wave-function of the first quantized theory and fields
in the second quantized theory, the fields in string field theory are represented as
states in the SCFT.  In the full BV quantized theory there are two sets of string fields:
$|\Psi\rangle\in\wh\HH_T$ and $|\wt\Psi\rangle\in\wt\HH_T$ without any further restriction.
The action is given by\footnote{We shall work in the convention in which the path integral
is carried out with the weight factor $e^S$. If we want to change this, {\it e.g.} use the
weight factor $e^{-S}$ or $e^{iS}$, we can achieve this by replacing $g_S^2$ by
$-g_S^2$ or $-ig_S^2$ in all subsequent formul\ae.
}
\be\label{eactorg}
S={1\over g_S^2} \left[ -{1\over 2} \langle \wt\Psi | c_0^- Q_B \GG |\wt\Psi\rangle
+ \langle \wt\Psi | c_0^- Q_B |\Psi\rangle + \sum_{n=1}^\infty {1\over n!}
\cL \Psi^n\cR \right]
\, ,
\ee
where $g_S$ is the string coupling, and $\cL A_1\cdots A_N\cR$ is a multilinear function
of $|A_1\rangle,\cdots |A_N\rangle\in\wh\HH_T$, constructed by first computing the
correlation functions of these
vertex operators together with PCO's and other ghost insertions
on Riemann surfaces, and then integrating the result 
over certain {\it subspaces} of the moduli space of Riemann surfaces with punctures.  
The string fields $|\Psi\rangle$ and
$|\wt\Psi\rangle$ are grassmann even i.e.\ the coefficient of a grassmann even (odd) 
basis state in the expansion of the string field is grassmann even (odd).  
In order to avoid cluttering up various formul\ae\ due to sign factors, 
we shall multiply the grassmann odd vertex operators of the CFT
by grassmann odd c-numbers and 
work with states $A_i$ 
which are all grassmann even. Whenever needed we can always strip
off these factors from both sides of an equation 
at the cost of picking up appropriate signs.
With this convention $\cL A_1\cdots A_N\cR$ 
is symmetric under exchange of external states and
satisfies the relation, 
\ben \label{evertex}
&&\sum_{i=1}^N \cL A_1\cdots A_{i-1} (Q_B A_i)
A_{i+1} \cdots A_N\cR  \nonumber \\
&=& -  
{1\over 2} \sum_{\ell,k\ge 0\atop \ell+k=N} \sum_{\{ i_a;a=1,
\cdots \ell\} , \{ j_b;b=1,\cdots k\} \atop
\{ i_a\} \cup \{ j_b\}  = \cL 1,\cdots N\}
}\cL A_{i_1} \cdots A_{i_\ell} \vp_s\cR  \cL \vp_r A_{j_1} \cdots A_{j_k}\cR 
\langle \vp_s^c | c_0^- \GG | \vp_r^c\rangle \nonumber \\
&& -   {1\over 2} \, g_S^2\, \cL A_1 \cdots A_N \vp_s \vp_r \cR \, \langle \vp_s^c | c_0^- \GG 
| \vp_r^c\rangle
\, .
\een
Using this relation one can show that the action \refb{eactorg} satisfies BV master
equation\cite{1508.05387} 
and can be quantized in the Siegel gauge $b_0^+|\Psi\rangle=0$, 
$b_0^+|\wt\Psi\rangle=0$. 

We shall now 
describe the Siegel gauge propagator\cite{1508.02481}. 
Since only the $\Psi$ field appears in the interaction, the relevant propagator  is
the $\Psi-\Psi$ propagator.
Instead of giving its
expression directly we shall describe it by its
operation of joining two Feynman diagrams.
Let us suppose that $f(A_1,\cdots A_m, \vp_s)$ denotes the contribution 
to the off-shell amplitude\footnote{Throughout this paper we shall mean by off-shell amplitude
the truncated Green's function where the tree level
propagators for external states are dropped.}
from a
specific Feynman diagram with external states $A_1,\cdots A_m, \vp_s\in\wh\HH_T$ 
and
$g(B_1,\cdots B_n, \vp_r)$ denotes the contribution from another
Feynman diagram with external states $B_1,\cdots B_n, \vp_r\in\wh\HH_T$. 
In both we 
use a normalization 
such that the contribution to $f(A_1,\cdots A_m, \vp_s)$ from the elementary vertex
is just $\cL A_1\cdots A_m\vp_s\cR$, and similarly for $g$. Generically
$f$ and
$g$ have no symmetry property since we are not considering sum over all diagrams.
Now we can construct another Feynman diagram with external states 
$A_1,\cdots A_m, B_1,\cdots B_n$ by joining $\vp_s$ and $\vp_r$ by a propagator, and
summing over $s$ and $r$. Its contribution is given by\footnote{The
expression for the propagator is somewhat different from the standard one (see
{\it e.g.} \cite{1508.02481})
where instead of a $c_0^-$ the propagator contained a $b_0^-$. This difference can be
traced to the inclusion of the $c_0^-$ in the normalization \refb{ebas} of the
basis states.}
\be \label{efa1}
-f(A_1,\cdots A_m, \vp_s) \, g(B_1,\cdots B_n, \vp_r) \,
\langle \vp_s^c | c_0^- b_0^{+} (L_0^+)^{-1} \GG | \vp_r^c\rangle\, .
\ee
Note that $f$ and/or $g$ may have odd grassmann parity from the grassmann odd
numbers
hidden inside the $A_i$'s, so one should be careful about their relative positioning.
Similarly if $f(A_1,\cdots A_n, \vp_s, \vp_r)$ denotes a Feynman diagram
with external states $A_1,\cdots A_n,\vp_s,\vp_r$ and if we consider a new
Feynman diagram obtained by joining $\vp_s$ and $\vp_r$ by a propagator and
summing over all choices of $\vp_s$, $\vp_r$, the new Feynman diagram is
given by
\be \label{efa2}
-  {1\over 2}  g_S^2\, f(A_1,\cdots A_m, \vp_s,\vp_r) 
\langle \vp_s^c | c_0^- b_0^{+} (L_0^+)^{-1} \GG | \vp_r^c\rangle\, .
\ee
The power of $g_S$ reflects that this operation increases the number of loops 
in the diagram by 1. The factor of 1/2 is a combinatorial factor.

Using this propagator and the elementary vertices encoded in the 
interaction term in \refb{eactorg} 
one can compute off-shell amplitudes of string field theory. 
Using standard manipulations these can be expressed as integrals 
over the moduli space of Riemann 
surfaces, with the integrand given by the
correlation function of vertex operators of external states, PCO's and ghost fields
on the Riemann surface.
The correlation function depends on the choice of 
local holomorphic
coordinates around the punctures and the PCO locations. 
String field theory provides us with these data.  It also gives us a cell decomposition of the moduli space such that
the contribution from each cell can be identified with a Feynman diagram of the string
field theory. These choices are not unique, but are tightly constrained, and different
choices give equivalent string field theories related by field redefinition\cite{9301097}. 
Given a string field theory we can not only
define the full off-shell amplitude, but consider other quantities commonly used in
quantum field theories {\it e.g.} 1PI amplitudes obtained
by summing over certain subset of diagrams. In the language of Riemann surface,
this means that we only integrate over certain subspaces of the moduli space.
We shall now describe some properties of these amplitudes.

Let $G(A_1,\cdots A_N)$ be the full off-shell truncated 
Green's function with external
states $A_1,\cdots A_N$, 
obtained by summing over all Feynman diagrams with
external states $A_1,\cdots A_N$, but dropping the tree level propagators of the
external states. We impose Siegel gauge condition on the internal states,
but take the external states $A_1,\cdots A_N$ to be arbitrary elements of 
$\wh \HH_T$.
Then $G(A_1,\cdots A_N)$ will be given by a sum of terms, each of which is given
by a product of the propagators and vertices. Let us now
consider the combination
$\sum_{i=1}^N G(A_1,\cdots A_{i-1}, Q_B A_i, \break A_{i+1},
\cdots A_N)$. Since each $A_i$ must come from some vertex
$\cL\cdots \cR$ in a given Feynman diagram,
the sum over $i$ can be organized into subsets, where in a given subset $Q_B$ acts
on different external states of the same vertex. This can then be simplified
using \refb{evertex}. This gives\footnote{To the best of our knowledge 
this form of the Ward identity has not been written down before even for closed bosonic
string field theory. While its consequence, described in \refb{efullG}, has a standard form,
we should add that this is not a convenient form for analyzing properties of on-shell
amplitudes. The Green's function $G$ has self energy insertions on the external legs and
therefore diverges on-shell. One needs to work with its cousin $\Gamma$ described in
\cite{1508.02481} where the full external propagators are removed and work with the
Ward identities satisfied by $\Gamma$ which take a different form. 
If there are tadpoles of
light fields then $G$ will be divergent even for off-shell external states and
\refb{egid}, \refb{efullG} are formal. In this case one has
to first construct the 1PI effective action, find its extremum and expand the action around this extremum
to compute the off-shell amplitudes\cite{1508.02481}. 
Nevertheless we present this analysis here since
in \S\ref{s3} we shall describe a similar analysis where all relevant quantities will be
manifestly free from divergences. \label{fo1}} 
\ben \label{egid} 
&&\hskip .2in \sum_{i=1}^N  G(A_1,\cdots A_{i-1}, Q_B A_i,
A_{i+1} \cdots A_N)  \nonumber \\
&& = -  
{1\over 2} \sum_{\ell,k\ge 0\atop \ell+k=N} \sum_{\{ i_a;a=1,
\cdots \ell\} , \{ j_b;b=1,\cdots k\} \atop
\{ i_a\} \cup \{ j_b\}  = \cL 1,\cdots N\}
}G( A_{i_1}, \cdots A_{i_\ell} ,\vp_s)  G( \vp_r, A_{j_1} \cdots A_{j_k})
\langle \vp_s^c | c_0^- \GG | \vp_r^c\rangle \nonumber \\
&& -   {1\over 2} \, g_S^2\, G(A_1, \cdots A_N ,\vp_s, \vp_r) \, \langle \vp_s^c | c_0^- \GG 
| \vp_r^c\rangle \nonumber \\
&& -  
{1\over 2} \sum_{\ell,k\ge 0\atop \ell+k=N} \sum_{\{ i_a;a=1,
\cdots \ell\} , \{ j_b;b=1,\cdots k\} \atop
\{ i_a\} \cup \{ j_b\}  = \cL 1,\cdots N\}
}\bigg[ - G( A_{i_1}, \cdots A_{i_\ell} ,Q_B\vp_s)  G( \vp_r, A_{j_1} \cdots A_{j_k})
\nonumber \\
&& \hskip 1in - (-1)^{\gamma_s}
G( A_{i_1}, \cdots A_{i_\ell} ,\vp_s)  G( Q_B\vp_r, A_{j_1} \cdots A_{j_k}) \bigg]
\langle \vp_s^c | c_0^- b_0^{+} (L_0^+)^{-1} \GG | \vp_r^c\rangle
\nonumber \\
&& -   {g_S^2\over 2} \bigg[-G(A_1, \cdots A_N ,Q_B\vp_s, \vp_r) 
- (-1)^{\gamma_s} G(A_1, \cdots A_N ,\vp_s, Q_B\vp_r)\bigg]\, \langle \vp_s^c | c_0^- 
b_0^{+} (L_0^+)^{-1}  \GG 
| \vp_r^c\rangle
\, . \nonumber \\ 
\een 
The first two terms on the right hand side 
represent the contribution from the right hand side of \refb{evertex} when we
use \refb{evertex} to simplify the contribution from  
individual vertices of the Feynman diagram.
The other two terms on the right hand side come
from the fact that while using \refb{evertex} for a given vertex, 
we have to subtract the terms where $Q_B$ acts on the
legs of the vertex connected to internal propagators
since on the left
hand side of \refb{egid} $Q_B$ only acts on the external states. The third term represents the
contribution from a propagator that connects two disjoint Feynman diagrams,
whereas the last term represents the contribution from a propagator that
connects two external lines of a connected Feynman diagram. The overall minus signs
in front of the third and the fourth terms come from having to move these from the left
hand side of the equation, where they appear naturally, to the right hand side.
The minus signs inside the square brackets
come from the ones on the right hand sides of \refb{efa1} and \refb{efa2}. The
$(-1)^{\gamma_s}$ factors arise from having to move $Q_B$ through $\vp_s$.
In
the third term, we have included a factor of 1/2 to 
compensate for the double counting
associated with the $\{i_a\}\leftrightarrow \{j_b\}$ exchange. 
The 1/2 in the last factor arises from the right hand side of \refb{efa2}.

Using the completeness relation \refb{ecom} we can now move $Q_B$ inside the
matrix element $\langle \vp_s^c | c_0^- 
b_0^{+} (L_0^+)^{-1}  \GG 
| \vp_r^c\rangle$ in the third and the fourth terms, {\it e.g.} we have 
\be
Q_B |\vp_s\rangle \langle \vp_s^c | c_0^- 
b_0^{+} (L_0^+)^{-1}  \GG 
| \vp_r^c\rangle = Q_B b_0^{+} (L_0^+)^{-1}  \GG 
| \vp_r^c\rangle = |\vp_s\rangle \langle \vp_s^c | c_0^- 
Q_B b_0^{+} (L_0^+)^{-1}  \GG 
| \vp_r^c\rangle \, , 
\ee
and 
\ben
&& (-1)^{\gamma_s} Q_B |\vp_r\rangle \langle \vp_s^c | c_0^- 
b_0^{+} (L_0^+)^{-1}  \GG 
| \vp_r^c\rangle =  Q_B |\vp_r\rangle \langle \vp_r^c 
| c_0^-  b_0^{+} (L_0^+)^{-1}  \GG 
| \vp_s^c\rangle 
=  Q_B 
b_0^{+} (L_0^+)^{-1}  \GG 
| \vp_s^c\rangle \nonumber \\ &=&
 |\vp_r\rangle \langle \vp_r^c | c_0^- Q_B 
b_0^{+} (L_0^+)^{-1}  \GG 
| \vp_s^c\rangle  = |\vp_r\rangle \langle \vp_s^c | c_0^-  
b_0^{+} Q_B (L_0^+)^{-1}  \GG 
| \vp_r^c\rangle
\, .
\een
Using the relations
$Q_Bb_0^++b_0^+ Q_B=L_0^+$ and \refb{egrel} one can now
show that on the right hand side of  \refb{egid} the third term
cancels the first term and the fourth term cancels the second
term. This gives us the Ward identity for the 
off-shell Green's function
\be \label{efullG}
\sum_{i=1}^N  G(A_1,\cdots A_{i-1}, Q_B A_i,
A_{i+1} \cdots A_N) =0\, .
\ee

As a simple example we can consider the tree level four point function, given by the
sum of s, t and u-channel diagrams and the four point vertex. As a direct 
consequence
of \refb{evertex}, the contribution of
the 4-point vertex to the right hand side of \refb{egid} will be given by
\ben \label{eexx1}
&& -G(A_1,A_2, \vp_s) G(\vp_r, A_3,A_4)\langle \vp_s^c | c_0^- \GG | \vp_r^c\rangle
-G(A_1,A_3, \vp_s) G(\vp_r, A_2,A_4)\langle \vp_s^c | c_0^- \GG | \vp_r^c\rangle
\nonumber \\
&& -G(A_1,A_4, \vp_s) G(\vp_r, A_2,A_3)]
\langle \vp_s^c | c_0^- \GG | \vp_r^c\rangle\, . 
\een
The contribution from the s-channel diagram will be given by 
\be \label{eexx2}
(G(A_1,A_2, Q_B\vp_s) G(\vp_r, A_3,A_4)+ (-1)^{\gamma_s} 
G(A_1,A_2, \vp_s) G(Q_B\vp_r, A_3,A_4)) \langle \vp_s^c | 
c_0^- b_0^{+} (L_0^+)^{-1} \GG | \vp_r^c\rangle\, .
\ee
After moving $Q_B$ inside the matrix element $\langle \vp_s^c | 
c_0^- b_0^{+} (L_0^+)^{-1} \GG | \vp_r^c\rangle$ using the completeness relation, and
using $Q_B b_0^++b_0^+Q_B=L_0^+$, 
this cancels the first term in \refb{eexx1}. Similarly the contribution from the t and u-channel
diagrams cancel the second and third terms in \refb{eexx1}.

An identity similar  to \refb{efullG} 
can be derived for the 1PI amplitudes. Let 
$\{A_1\cdots A_n\}$ denote the
1PI amplitude
of the external states $A_1,\cdots A_n$. 
This will
satisfy an identity similar to \refb{egid} with $G(A_1,\cdots A_n)$ replaced by
$\{A_1\cdots A_n\}$, and without the third term on the right hand side of 
\refb{egid}. This is due to the fact that by definition, 1PI amplitudes do not 
include sum over Feynman diagrams in which a single propagator connects two
other Feynman diagrams. Therefore the first term on the right hand side remains
uncanceled and we arrive at
the identity:
\ben\label{e1pi}
&&\sum_{i=1}^N \{A_1\cdots A_{i-1} (Q_B A_i)
A_{i+1} \cdots A_N\} \nonumber \\
&=& -  
{1\over 2} \sum_{\ell,k\ge 0\atop \ell+k=N} \sum_{\{i_a;a=1,\cdots \ell\}, \{j_b;b=1,\cdots k\}\atop
\{i_a\}\cup \{j_b\} = \{1,\cdots N\}
}\{A_{i_1} \cdots A_{i_\ell} \vp_s\} \{\vp_r A_{j_1} \cdots A_{j_k}\}
\langle \vp_s^c | c_0^- \GG | \vp_r^c\rangle\, .
\een

\sectiono{The effective action of light fields} \label{s3}

In this section we shall analyze
the  effective action for light fields obtained by integrating out the  heavy
fields. By definition, the $L_0^+-\alpha' k^2/2$ eigenvalue 
vanishes for the light states. We shall denote by $P$ the
projection operator into light states. $P$ satisfies\footnote{While 
$P$ projects to states of mass level zero, it keeps
all the momentum modes of the light fields. Therefore at this stage it will
be premature to claim that integrating
out modes outside the $P$ invariant 
subspace will lead to the Wilsonian
effective action. We shall see in the next section that by adjusting a parameter
in string
field theory we can effectively integrate out the high momentum modes of the light fields.}
\be \label{epprop}
[P,b_0^\pm]=0, \quad [P,L_0^\pm]=0, \quad [P, \GG]=0, \quad [P,Q_B]=0\, .
\ee
Consider a set of light off-shell states $a_1,\cdots a_N$. 
We denote by $\cL a_1\cdots a_N\cR_e$ the total contribution to
the amplitude with external states $a_1,\cdots a_N$ 
from {\it all} the Feynman diagrams of superstring field theory, but with the
propagator factors appearing in \refb{efa1}, \refb{efa2} replaced by  
$\langle \vp_s^c | c_0^- b_0^{+} (L_0^+)^{-1} \GG (1-P) | \vp_r^c\rangle$.
This removes the contributions of light fields from the
propagator. Therefore $\cL a_1\cdots a_N\cR_e$ can be regarded as the contribution 
to the off-shell amplitude due to the elementary
$N$-point vertex of the effective theory, obtained by 
integrating out the heavy fields.\footnote{Even in the presence of tadpoles of 
light fields, these
amplitudes do not suffer from divergences of the kind mentioned in footnote \ref{fo1}.}
We can now repeat the argument leading to \refb{egid} with 
$G(\cdots)$ replaced by $\cL \cdots\cR_e$.
On the left hand side of \refb{egid} and the first two terms on the right hand side
of \refb{egid} we simply replace $G(\cdots)$ by $\cL\cdots\cR_e$, 
but in the last two terms 
of \refb{egid} the propagator factors will now have additional insertions of $(1-P)$ 
since this is the propagator used in the definition of $\cL\cdots\cR_e$. 
This gives \eject 
\ben \label{ewid}
&&\sum_{i=1}^N  \cL a_1\cdots a_{i-1} (Q_B a_i)
a_{i+1} \cdots a_N\cR_e  \nonumber \\
&=& -  
{1\over 2} \sum_{\ell,k\ge 0\atop \ell+k=N} \sum_{\{ i_a;a=1,
\cdots \ell\} , \{ j_b;b=1,\cdots k\} \atop
\{ i_a\} \cup \{ j_b\}  = \cL 1,\cdots N\}
}\cL a_{i_1} \cdots a_{i_\ell} \vp_s\cR_e  \cL \vp_r a_{j_1} \cdots a_{j_k}\cR_e
\langle \vp_s^c | c_0^- \GG | \vp_r^c\rangle \nonumber \\
&& -  {1\over 2} \, g_S^2 \cL a_1 \cdots a_N \vp_s \vp_r\cR_e \, \langle \vp_s^c | c_0^- \GG 
| \vp_r^c\rangle \nonumber \\
&& -  
{1\over 2} \sum_{\ell,k\ge 0\atop \ell+k=N} \sum_{\{ i_a;a=1,
\cdots \ell\} , \{ j_b;b=1,\cdots k\} \atop
\{ i_a\} \cup \{ j_b\}  = \cL 1,\cdots N\}
}\bigg[ -\cL a_{i_1} \cdots a_{i_\ell}  (Q_B\vp_s)\cR_e  \cL \vp_r a_{j_1} \cdots a_{j_k}\cR_e
\nonumber \\
&& \hskip .5in - (-1)^{\gamma_s}
\{ a_{i_1} \cdots a_{i_\ell} \vp_s\cR_e  \cL (Q_B\vp_r) a_{j_1} \cdots a_{j_k}\cR_e \bigg]
\langle \vp_s^c | c_0^- b_0^{+} (L_0^+)^{-1} \GG (1-P)| \vp_r^c\rangle
\nonumber \\
&& -  {1\over 2} \, g_S^2 \bigg[-\cL a_1 \cdots a_N (Q_B\vp_s) \vp_r\cR_e
- (-1)^{\gamma_s} \cL a_1 \cdots a_N \vp_s (Q_B\vp_r)\cR_e \bigg]\, 
\nonumber \\ && 
\hskip 1in \langle \vp_s^c | c_0^- 
b_0^{+} (L_0^+)^{-1}  \GG (1-P)
| \vp_r^c\rangle
\, . 
\een  
\noindent 
Now the third and the fourth terms on the right hand side cancel the first and
the second terms only partially, leaving behind terms proportional to
$\langle \vp_s^c | c_0^- \GG 
\, P | \vp_r^c\rangle$:
\ben \label{ewidmid}
&&\sum_{i=1}^N  \cL a_1\cdots a_{i-1} (Q_B a_i)
a_{i+1} \cdots a_N\cR_e  \nonumber \\
&=& -  
{1\over 2} \sum_{\ell,k\ge 0\atop \ell+k=N} \sum_{\{ i_a;a=1,
\cdots \ell\} , \{ j_b;b=1,\cdots k\} \atop
\{ i_a\} \cup \{ j_b\}  = \cL 1,\cdots N\}
}\cL a_{i_1} \cdots a_{i_\ell} \vp_s\cR_e  \cL \vp_r a_{j_1} \cdots a_{j_k}\cR_e
\langle \vp_s^c | c_0^- \GG \, P| \vp_r^c\rangle \nonumber \\
&& -  {1\over 2} \, g_S^2 \cL a_1 \cdots a_N \vp_s \vp_r\cR_e \, \langle \vp_s^c | c_0^- 
\GG\, P 
| \vp_r^c\rangle\, .
\een
If we denote by $\{|\chi_\alpha\rangle\}$ and 
$\{|\chi_\alpha^c\rangle\}$
the basis states  in $P\wh\HH_T$
and $P\wt\HH_T$ respectively, satisfying
\be \label{echic}
\langle\chi_\alpha|c_0^-|\chi^c_\beta\rangle = \delta_{\alpha\beta}, \quad
\langle\chi^c_\beta|c_0^-|\chi_\alpha\rangle = \delta_{\alpha\beta},
\ee 
then we can express \refb{ewidmid} as 
\ben \label{ewidfin}
&&\sum_{i=1}^N  \cL a_1\cdots a_{i-1} (Q_B a_i)
a_{i+1} \cdots a_N\cR_e  \nonumber \\
&=& -  
{1\over 2} \sum_{\ell,k\ge 0\atop \ell+k=N} \sum_{\{ i_a;a=1,
\cdots \ell\} , \{ j_b;b=1,\cdots k\} \atop
\{ i_a\} \cup \{ j_b\}  = \cL 1,\cdots N\}
}\cL a_{i_1} \cdots a_{i_\ell} \chi_\alpha\cR_e  \cL \chi_\beta a_{j_1} \cdots a_{j_k}\cR_e
\langle \chi_\alpha^c | c_0^- \GG | \chi_\beta^c\rangle \nonumber \\
&& -  {1\over 2} \, g_S^2\, \cL a_1 \cdots a_N \chi_\alpha\chi_\beta\cR_e \, \langle 
\chi_\alpha^c | c_0^- \GG 
| \chi_\beta^c\rangle\, .
\een 
\noindent This may be considered as the key technical result of this paper.

Given the identity \refb{ewidfin}
one can now construct the string field theory action satisfying BV
master equation in a straightforward manner following the procedure described in
\cite{1508.05387} and reviewed in \S\ref{s2}. 
We introduce two sets of grassmann even string fields, 
$\Phi\in P\wh\HH_T$ and $\wt\Phi\in P\wt\HH_T$, i.e.\ 
both containing only light 
states.
The effective master action is given by
\be\label{emaster}
S_e={1\over g_S^2} \left[ -{1\over 2} \langle \wt\Phi | c_0^- Q_B \GG |\wt\Phi\rangle
+ \langle \wt\Phi | c_0^- Q_B |\Phi\rangle + \sum_{n=1}^\infty {1\over n!}
\cL \Phi^n\cR_e\right]
\, .
\ee
Even though we used Siegel gauge
to integrate out the heavy fields, there is no restriction on $|\Phi\rangle$ and 
$|\wt\Phi\rangle$ to satisfy the Siegel gauge condition. This is in the spirit
of the BV formalism where the master action is constructed {\it before gauge
fixing}.
In order to show that \refb{emaster} satisfies the BV master
equation we proceed as follows\cite{1508.05387}:
\begin{enumerate}
\item We define $\wh\HH_+$
and $\wt\HH_+$
to be the subspaces of $\wh\HH_T$ and $\wt\HH_T$ respectively containing states of
ghost numbers $\ge 3$, and  $\wh\HH_-$
and $\wt\HH_-$ to be the subspaces of 
$\wh\HH_T$ and $\wt\HH_T$ respectively containing 
states of
ghost numbers $\le 2$. We organize the basis states $\{|\chi_\alpha\rangle\}$ 
into $\{|\wh\chi^-_\alpha\rangle\}$ and $\{|\wh\chi_+^\alpha\rangle\}$ of $P\wh\HH_-$ and
$P\wh\HH_+$ respectively, 
and the basis states
$\{|\chi_\alpha^c\rangle\}$ into  basis states 
$\{|\wt\chi^-_\alpha\rangle\}$  and $\{|\wt \chi_+^\alpha\rangle\}$ of
$P\wt\HH_-$, 
and $P\wt\HH_+$ respectively, satisfying orthonormality conditions\footnote{By an abuse
of notation we are using the same indices $\alpha$ to label the new basis even though
the label runs over a smaller set for the new basis.}
\be \label{einner}
\langle \wh\chi^-_\alpha|c_0^- |\wt \chi_+^\beta \rangle = \delta_\alpha{}^\beta=
\langle \wt \chi_+^\beta|c_0^- |\wh\chi^-_\alpha\rangle, \quad 
\langle \wt\chi^-_\alpha|c_0^- |\wh \chi_+^\beta \rangle = \delta_\alpha{}^\beta=
\langle \wh \chi_+^\beta|c_0^- |\wt\chi^-_\alpha\rangle\, ,
\ee
and the completeness relations
\be \label{eortho}
\sum_\alpha |\wh\chi^-_\alpha\rangle \langle \wt \chi_+^\alpha|c_0^-
+ \sum_\alpha |\wh\chi^\alpha_+\rangle\langle \wt \chi_\alpha^-|c_0^- ={\bf 1}, 
\quad 
\sum_\alpha |\wt\chi^-_\alpha\rangle \langle \wh \chi_+^\alpha|c_0^-
+ \sum_\alpha |\wt\chi^\alpha_+\rangle\langle \wh \chi_\alpha^-|c_0^- ={\bf 1}\, ,
\ee
acting on states in $P\wh\HH_T$ and $P\wt\HH_T$ respectively.
\item
The light string fields $|\Phi\rangle$ and $|\wt\Phi\rangle$ are expanded as
\ben \label{ephiexpan}
|\wt\Phi\rangle &=& \sum_\alpha |\wt\chi^-_\alpha\rangle \wt\phi^\alpha
+\sum_\alpha (-1)^{\gamma_\alpha^*+1} |\wt\chi_+^\alpha\rangle \phi_\alpha^*\, , \nonumber \\
|\Phi\rangle -{1\over 2} 
\GG|\wt\Phi\rangle &=& \sum_\alpha |\wh\chi^-_\alpha\rangle \phi^\alpha
+ \sum_\alpha (-1)^{\wt \gamma_\alpha^*+1} |\wh\chi_+^\alpha\rangle \wt\phi_\alpha^* 
\, .
\een
Here $\gamma^*_\alpha$, $\gamma_\alpha$, $\wt \gamma^*_\alpha$ 
and $\wt \gamma_\alpha$ label the grassmann
parities of $\phi^*_\alpha$, $\phi^\alpha$, $\wt \phi^*_\alpha$ and $\wt\phi^\alpha$
respectively. 
We shall identify the variables 
$\{\phi^\alpha, \wt\phi^\alpha\}$ as `fields' and the
variables $\{\phi^*_\alpha, \wt\phi^*_\alpha\}$ as the conjugate `anti-fields' in the BV
quantization of the theory. 
\item Given two functions $F$ and $G$ of all the fields and anti-fields,
we now define their anti-bracket as:
\be \label{eantib}
\{F, G\}= {\p_R F\over \p \phi^\alpha} \, {\p_L G\over \p\phi^*_\alpha}
+ 
{\p_R F\over \p \wt\phi^\alpha} \, {\p_L G\over \p\wt\phi^*_\alpha}
- {\p_R F\over \p \phi^*_\alpha} \, {\p_L G\over \p\phi^\alpha}
-
{\p_R F\over \p \wt\phi^*_\alpha} \, {\p_L G\over \delta\wt\phi^\alpha}\, ,
\ee
where the subscripts $R$ and $L$ of $\p$ denote left and right derivatives 
respectively. We also define
\be \label{edefDelta}
\Delta F \equiv {\p_R\over \p\phi^\alpha} {\p_L F\over \p \phi^*_\alpha}
+ {\p_R\over \p\wt\phi^\alpha} {\p_L F\over \p \wt\phi^*_\alpha}\, .
\ee
\item Using \refb{eortho}, \refb{ephiexpan} and \refb{eantib} one gets, after some algebra,
\be\label{ess}
g_S^4 \{S_e, S_e\} = - 2  \sum_{n} {1\over (n-1)!} \cL \Phi^{n-1} Q_B\Phi\cR_e
- \sum_{m,n} {1\over m! n!} \cL \chi_\beta \Phi^{m}\cR_e \cL \chi_\alpha
\Phi^{n}\cR_e \, \langle \chi_\beta^c |c_0^- \GG | \chi_\alpha^c\rangle \, . 
\ee
Here $|\chi_\alpha\rangle$'s denote the original choice of basis states in $\wh\HH_T$
before splitting it into $\wh\HH_\pm$.
\item On the other hand using \refb{eortho}, \refb{ephiexpan} and 
\refb{edefDelta} we get
\be \label{edde}
\Delta S_e = - {1\over 2\, g_S^2}\sum_n {1\over n!} \cL \Phi^n \chi_\beta \chi_\alpha\cR_e
\langle \chi_\beta^c  | c_0^- \GG |\chi^c_\alpha\rangle \, .
\ee
\item 
Using the identity \refb{ewidfin}, and eqs.\refb{ess}, \refb{edde}
one can show that the action $S_e$ given in  \refb{emaster}
satisfies the quantum BV master equation
\be \label{eqmaster}
{1\over 2} \{S_e, S_e\} + \Delta S_e = 0\, .
\ee
\end{enumerate}

One can choose Siegel gauge $b_0^+|\Phi\rangle=0=b_0^+|\wt\Phi\rangle$ and
derive the Feynman rules in a straightforward manner following a procedure identical to the
one used for the full string field theory. 
One finds that the field $\Phi$ describes an interacting field, while the degrees
of freedom associated with $\wt\Phi$ describe decoupled free fields.
By analyzing the amplitudes with external $\Phi$ legs one
arrives at the conclusion that when two Feynman diagrams are joined by a $\Phi$
propagator one has an expression analogous to \refb{efa1}
\be \label{efa1mod}
-f(a_1,\cdots a_m, \chi_\alpha) \, g(b_1,\cdots b_n, \chi_\beta) \,
\langle \chi_\alpha^c | c_0^- b_0^{+} (L_0^+)^{-1} \GG | \chi_\beta^c\rangle\, .
\ee
On the other hand when two legs of a connected Feynman diagram are joined by a $\Phi$
propagator, we have the analog of \refb{efa2}
\be \label{efa2mod}
- {1\over 2}\, g_S^2\, f(a_1,\cdots a_m, \chi_\alpha,\chi_\beta) 
\langle \chi_\alpha^c | c_0^- b_0^{+} (L_0^+)^{-1} \GG | \chi_\beta^c\rangle\, .
\ee
By writing down an analog of \refb{egid} in the effective field theory one can show that
the full amplitude constructed by summing over all Feynman diagrams
of light fields with $\cL a_1\cdots a_N\cR_e$ as elementary vertices
satisfies the same identity as \refb{efullG}.
This is of course a reflection of the fact that the full amplitude constructed from
the effective action and the original string field theory action are
identical.\footnote{As mentioned in footnote \ref{fo1}, if there are tadpoles of light
states then the full amplitude is divergent. In such cases one has to first construct the
1PI 
effective action, find the extremum of this action and then compute the Green's functions
by expanding the action around the new background.}

We can also introduce the notion of 1PI amplitudes 
$\{a_1\cdots a_N\}_e$ built from the elementary 
vertices described above. By proceeding in the same way as from \refb{evertex}
to \refb{e1pi}, one can show that $\{a_1\cdots a_N\}_e$ satisfies the identity
\ben
&&\sum_{i=1}^N  \{ a_1\cdots a_{i-1} (Q_B a_i)
a_{i+1} \cdots a_N\}_e  \nonumber \\
&=& -  
{1\over 2} \sum_{\ell,k\ge 0\atop \ell+k=N} \sum_{\{ i_a;a=1,
\cdots \ell\} , \{ j_b;b=1,\cdots k\} \atop
\{ i_a\} \cup \{ j_b\}  = \{ 1,\cdots N\}
}\{ a_{i_1} \cdots a_{i_\ell} \chi_\alpha\}_e  \{ \chi_\beta a_{j_1} \cdots a_{j_k}\}_e
\langle \chi_\beta^c | c_0^- \GG | \chi_\alpha^c\rangle \, . \nonumber \\
\een
Using this one can construct 1PI action for light fields:
\be 
S_e^{1PI} = {1\over g_S^2} \left[ -{1\over 2} \langle \wt\Phi | c_0^- Q_B \GG |\wt\Phi\rangle
+ \langle \wt\Phi | c_0^- Q_B |\Phi\rangle + \sum_{n=1}^\infty {1\over n!}
\{ \Phi^n\}_e\right]
\, .
\ee
This satisfies the classical master equation
\be
\{ S_e^{1PI}, S_e^{1PI}\}=0\, .
\ee
Furthermore by restricting the fields $|\Phi\rangle$ and $|\wt\Phi\rangle$ to  
carry ghost number 2, 
we arrive at the gauge invariant
1PI action of space-time matter fields with gauge transformation laws
\be
\delta |\wt\Phi\rangle = Q_B|\wt\Lambda\rangle +\sum_n {1\over n!} [\Phi^n\Lambda]_e\, ,
\quad \delta |\Phi\rangle = Q_B|\Lambda\rangle +\sum_n {1\over n!} \GG\, 
[\Phi^n\Lambda]_e
\, .
\ee
Here the gauge transformation parameters 
$|\Lambda\rangle$ and $|\wt\Lambda\rangle$ are states
of $P\, \wh\HH_T$ and
$P\, \wt\HH_T$ respectively, carrying ghost number 1.
For $|a_i\rangle\in P\wh\HH_T$,
$[a_1\cdots a_N]_e$ describes a state in $P\, \wt\HH_T$ defined via
\be
\langle a_0|c_0^- | [a_1\cdots a_N]_e\rangle = \{a_0a_1\cdots a_N\}_e\, ,
\quad \forall |a_0\rangle \in P\wh\HH_T\, .
\ee

The 1PI action so constructed is useful for analyzing various properties of 
light fields. For example the quantum corrected vacuum can be found by identifying
the translationally invariant extremum of the action. The 
renormalized masses can be computed directly by
solving the linearized equations of motion around the extremum of the action.\footnote{Even
though in ten dimensions the masses of light states do not get 
renormalized due to gauge invariance, this is not always
true in lower dimensions, {\it e.g.} in the situation analyzed in \cite{1508.02481}.}

Before concluding this section, we shall give some physical insight
into the definition of the vertex $\cL a_1\cdots a_N\cR_e$. We have defined it as the
result of building up the off-shell amplitude of light states
by summing over Feynman diagrams of superstring
field theory after subtracting off the contribution due to light states from the
propagators. This would be
a complicated procedure involving piecing  together 
amplitudes of heavy states with the help of \refb{efa1}, \refb{efa2} with additional
insertion of $(1-P)$ factors in the propagators. 
A simpler definition involves proceeding from the other end, 
where we begin with the full off-shell  amplitude of light states and carry out
appropriate subtraction involving light intermediate states. 
A systematic procedure for doing this has been
described in appendix \ref{sb}. The full amplitude has simple 
interpretation as the integrals of 
correlation functions of vertex operators of light states over the full moduli space of Riemann
surfaces.  The subtraction terms also involve products of light state propagators and
correlation functions of off-shell vertex operators of light states over lower genus Riemann
surfaces. Therefore with this definition one never has to worry about computing off-shell
amplitudes of heavy states. The analysis nevertheless requires some data from string field
theory since the definition of the off-shell amplitude, as well as the association of 
subspaces of the moduli space with the Feynman diagrams, requires 
data from string field theory. 
We shall elaborate on this in \S\ref{s6}.
A similar procedure can be adapted for defining the 1PI vertices $\{a_1\cdots a_N\}_e$.
In this case the subtraction of the contribution of light states has to be applied
only on those propagators which connect two Feynman diagrams that are otherwise
disjoint. 

\sectiono{The effective action as Wilsonian effective action} \label{s4}

In this section we shall argue following \cite{br1,br2}
that the effective action derived in \S\ref{s3} is
actually a Wilsonian effective action in which the high momentum modes of the
light fields are also effectively integrated out. For this we shall first 
illustrate this mechanism in the context of a quantum field theory.

Consider an ordinary quantum field theory, possibly containing multiple fields of
different masses. A Feynman diagram of this theory is
given by the integral over  loop momenta of an integrand that is given by the product
of propagators and vertices. Let us represent the denominator factor of
each propagator as an integral over a Schwinger parameter as
follows:
\be
(k^2+m^2)^{-1} =\int_0^\infty ds \, e^{-s (k^2+m^2)}\, .
\ee
Now let us replace the right hand side by
\be \label{edecomp}
\int_\Lambda^\infty ds \, e^{-s(k^2+m^2)} + \int_0^\Lambda ds\, e^{-s(k^2+m^2)}
= e^{-\Lambda(k^2+m^2)} (k^2+m^2)^{-1} +\int_0^\Lambda ds\, e^{-s(k^2+m^2)}\, .
\ee
The first term contains a pole at $k^2+m^2=0$. Let us redefine the vertices
by absorbing a factor of $e^{-\Lambda(k^2+m^2)/2}$ in each of the two vertices
that the propagator connects to. If we do this for each propagator and pick the
first term in each propagator then the result will have an interpretation of Feynman
diagram in a new theory in which each vertex is scaled by a factor of 
$e^{-\Lambda \sum_i (k_i^2+m_i^2)/2}$, with the sum  running over all the external
legs of the vertex.  This still leaves us with the second term in \refb{edecomp} which needs
to be taken into account. Since this does not have any pole at $k^2+m^2=0$, its 
contribution may be represented by including a further additive contribution to the
vertices. This leads to a new set of vertices (and the standard propagator) which give
identical contribution to the S-matrix as the original vertices.

As a simple example we can consider a tree level four point amplitude corresponding
to the `$s$-channel diagram'. Representing the propagator as
in \refb{edecomp}, we can regard the contribution to the four point function 
from the second term in \refb{edecomp} as coming from a new additive
contribution to the
four point vertex. On the other hand the $e^{-\Lambda(k^2+m^2)}$ factor in the first
term can be absorbed into a multiplicative factor of $e^{-\Lambda(k^2+m^2)/2}$ in each
of the two three point vertices.
This procedure can be carried out for all tree and loop amplitudes.

It is clear that this procedure removes some of the contribution from the composite
Feynman diagrams and transfers it to the definition of the elementary vertices. This
is precisely the notion of integrating out certain set of degrees of freedom to generate
a new but equivalent theory. It is also easy to see which degrees of freedom are being
integrated out. For this let us consider the case of massless fields and euclidean theory
so that $k^2$ is positive. In this case most of the contribution of the modes with
$k^2>>\Lambda^{-1}$ is in the second term in \refb{edecomp}
-- the contribution of these modes to the
first term is exponentially suppressed. Therefore including the contribution from the
second term in the vertices corresponds to integrating out the modes carrying 
momentum $>> \Lambda^{-1/2}$. 
This is precisely the notion of Wilsonian effective action.

Let us now return to string field theory. It turns out that string field theories automatically
come with the parameter $\Lambda$\cite{sonoda}. 
This corresponds to the freedom of rescaling the
local holomorphic coordinates around the punctures
by a constant. This rescaling automatically multiplies
the off-shell amplitudes (and hence also the elementary vertices) by a factor of
$e^{-\Lambda \sum_i (k_i^2+m_i^2)/2}$ where the $(k_i^2+m_i^2)$ factor arises from
the conformal weight of the vertex operator. However the rescaling of the local
holomorphic coordinates also needs to be accompanied by a change in the cell 
decomposition of the moduli space -- transferring some part of the moduli space corresponding
to composite Feynman diagrams to elementary vertices -- in order to preserve the 
relations \refb{evertex}. 
Therefore the operation of rescaling the local holomorphic coordinates
in the definition of off-shell amplitudes -- known in the string field theory literature as
the operation of adding stubs -- is precisely the operation of rearranging the
contribution from different Feynman diagrams in a quantum field theory using 
\refb{edecomp}. As argued before,
this operation 
is equivalent to integrating out the 
high momentum modes. Therefore we see that the effective action we have obtained 
automatically represents Wilsonian effective action in which by controlling the 
stub length we can adjust the energy scale above which the modes of the light
fields are integrated out. This also determines the effective UV cut-off of the theory,
encoded in the exponential suppression factor in the vertices.\footnote{Note however
that in a given scattering amplitude with some fixed total center of mass energy $E$
carried by the incoming particles, the loop energy integration contours veer away
from the imaginary axis by a distance of order $E$ in order to go around the poles
of the propagators\cite{1604.01783,1607.06500}. 
Since the vertices grow exponentially for real energy, 
a large stub length does not help us in reducing the range of loop energy
integration below $E$.
}

In the BV formalism the operation of adding stubs
is generated by a symplectic transformation of the fields and does not change the
physical theory\cite{9301097}. In particular this 
operation generates a field redefinition of
the 1PI effective action and leaves the S-matrix 
unchanged\cite{1411.7478,1501.00988}.

\sectiono{Discussion} \label{s6}

Construction of superstring field theory has two parts. The first part involves 
associating,
to each elementary vertex of the field theory with a certain number of external lines,
a certain subspace of the moduli space of punctured
Riemann surfaces, and specifying,  for each Riemann surface belonging to the subspace, 
a choice of local holomorphic coordinates around
the punctures and PCO locations.
The number of punctures match the number of external lines and the genus of the
Riemann surface determines the power of $g_S$ by which the contribution from
the vertex is multiplied. Generically a vertex with a given number of external lines
receives contribution from all genera.
The choice of this data  is not unique, but neither is this completely arbitrary as it
has to satisfy some stringent constraints that will be described below. 
Given this data one can draw all possible 
Feynman diagrams for a given amplitude, and for each of these Feynman diagrams 
associate a subspace of the moduli space of Riemann surfaces using the following
algorithm. When two elementary vertices are 
joined by a Feynman propagator to generate a new Feynman diagram as
in \refb{efa1}, this corresponds
to gluing the corresponding Riemann surfaces via the plumbing fixture relation
\be 
zw = e^{-s+i\theta}, \quad 0\le s<\infty, \quad 0\le\theta<2\pi\, ,
\ee
where $z$ and $w$ are the local holomorphic coordinates around the punctures
which are joined by the propagator.
This not only generates a family of Riemann surfaces corresponding to the Feynman
diagram but also specifies the local coordinates around the punctures and the PCO
locations on each member of this new family.\footnote{When two 
punctures associated with Ramond
sector states are joined this way,
one has an additional insertion of the zero mode of the PCO around one
of the punctures.} A similar interpretation can be given for the case where two lines
coming out of a single vertex are joined by a propagator as in \refb{efa2}. 
Repeating this procedure one can generate a family of 
Riemann surfaces corresponding to every Feynman diagram, together with choice of
local holomorphic coordinates around the punctures and PCO locations
for each member of this family. The consistency of the
original choice guarantees that the family of Riemann surfaces corresponding to all
Feynman diagrams covers the full moduli space of Riemann surfaces in a one to one
fashion and furthermore that the choice of local holomorphic coordinates around the
punctures and the PCO locations are continuous across the boundaries that separate
the subspaces of the moduli space associated with different Feynman diagrams.
 
In the second step one focusses on superstring theory in some specific background
associated with a choice of the superconformal field theory of matter and ghost fields
and computes the amplitude associated with a Feynman diagram 
by inserting the vertex operators at the
punctures using the predefined local holomorphic coordinates
and integrating the correlation function over the  corresponding subspace
of the moduli space of Riemann surfaces. Based on this one arrives at the definition 
of $\cL A_1\cdots A_N\cR$ and
the BV master action described in \S\ref{s2}. 

The result of our analysis can be 
interpreted by saying that these data can be used to construct other useful
quantities in string theory, {\it e.g.} Wilsonian effective action and 1PI effective action for
light fields. For the interaction vertices of the 
Wilsonian effective action we take the expression for the
full off-shell amplitude of light states and subtract from each Feynman diagram
the contribution to the propagator from the light states following
the procedure described in appendix \ref{sb}. For the 1PI effective action
we carry out a similar procedure, but perform
the subtraction only from the propagators connecting two
disjoint Feynman diagrams. For amplitudes with light fields as external states,
these actions contain the full information of the original string field theory.

The construction described in this paper can be generalized to other situations by replacing the
projection operator $P$ to light states by some other projection
operator satisfying the identities given in \refb{epprop}. In particular we could choose
$P$ to be a projection operator into another mass level, or into a set of mass levels.
Projecting into a fixed mass level may be useful, {\it e.g.} for computing the renormalized
masses of the states at that mass level, while projecting into a set of mass levels can be
useful for computing S-matrix elements where the external states are chosen from 
that set. This will automatically implement the procedure described in 
\cite{1311.1257,1401.7014,1411.7478,1501.00988} for integrating
out states at other mass levels while computing the renormalized mass at a given level.
Note however that if the theory has tadpoles of mass level zero fields requiring a shift
in the vacuum under quantum corrections then it is not possible to integrate out the
degrees of freedom associated with the mass level zero states.  The projection operator
$P$ must always include the mass level zero states so that the corresponding 1PI action
can be used to determine the correct vacuum before proceeding to compute the
renormalized masses. Even in the absence of tadpoles, an effective action in which the
fields of lower mass levels are integrated out will necessarily be complex, reflecting that
the scattering could produce the states that have been integrated out. For this reason, for
computing a given scattering amplitude, it is best to keep all the fields up to the mass level
that can be produced in the final state, and integrate out all fields above this mass level.
This will have the advantage of having to deal with only a finite number of fields, but
all the standard manipulations can be carried out with this action. For example the proof
of unitarity given in \cite{1607.08244} can be repeated with this formulation of the 
theory without any change.

Given the complexity of string field theory -- having infinite number of
fields --  one could wonder 
whether what we have learned from string field theory can be reformulated as
some suitable modification of the world-sheet description of scattering amplitudes. The
effective action just described can be regarded as  
the answer to this question. The data required for writing
down the action involves computation of correlation functions  on the
Riemann surface of only states below a certain mass level, 
with certain subtractions that also involves correlation functions of
the same states. Nevertheless the action so constructed is free from all divergences.
Therefore in this formalism all the divergences of the S-matrix appear as usual infrared 
divergences of a quantum field theory and can be dealt with using standard tools of
quantum field theory.

One could also consider other projection operators satisfying \refb{epprop}. 
As an example, consider toroidal compactification of type II 
string theory and take $P$ to be the projection operator on to states for 
which the left and right-moving oscillators are
forced to be in their GSO invariant ground state.
By integrating out all other fields following the procedure described
in this paper, one would arrive at the double field theory action envisaged in
\cite{0904.4664}. Such a field theory may be useful for describing scattering
of states in the kinematic regime in which the $P$ non-invariant states are
not produced in the final state.

It is tempting to conjecture that the form of the effective action
we have found extends beyond conventional string theory, {\it e.g.} the
Wilsonian effective action of the 
eleven dimensional M-theory may be given by a formula similar to \refb{emaster}
with the vertices satisfying the identities \refb{ewidfin}, 
even though they are not given
as integrals over the moduli space of Riemann surfaces. 
M-theory is known to exist\cite{9501068,9503124} but there is no
systematic method for computing scattering amplitudes beyond the leading
order due to the absence of a dimensionless expansion parameter. The
artificial dimensionless parameter given by the ratio of the cut-off scale and
the Planck scale could serve the purpose of an expansion parameter, with higher
loop contributions being suppressed by powers of this parameter as long as the
momenta of external states remain below the cut-off scale. From this perspective, 
construction of
M-theory reduces to the problem of finding solutions to \refb{ewidfin}
subject to the boundary condition at low momentum 
provided by the eleven dimensional
supergravity.

\bigskip

\noindent {\bf Acknowledgement:}
I wish to thank Barton Zwiebach for useful discussions and critical comments on
earlier versions of the manuscript.
This work was
supported in part by the 
DAE project 12-R\&D-HRI-5.02-0303 and J. C. Bose fellowship of 
the Department of Science and Technology, India.


\appendix

\sectiono{Construction of the interaction vertices of the effective action} \label{sb}

In this appendix we shall give a systematic procedure for constructing the interaction
vertices $\cL a_1\cdots a_N\cR_e$ that go into the definition of the Wilsonian
effective action. For this we shall need to introduce some notation. The original
superstring field theory provides us with a cell decomposition of the moduli
space  of Riemann surfaces with punctures, with every cell corresponding to a
particular Feynman diagram. It also specifies
the choice of local holomorphic coordinates at the punctures and the 
PCO locations for every Riemann surface.
This allows us to express the off-shell amplitude with external states $a_1,\cdots a_N$
as $\int I$
where the integral runs over the moduli space and the integrand $I$ is expressed in terms
of appropriate correlations functions on the Riemann surface. The contribution to
$\cL a_1\cdots a_N\cR_e$ will be given by a similar integral but with different integrand, taking
different forms inside different cells of the moduli space.
We shall
now describe how to construct the integrand.

First consider the cell  $C$ 
associated with the Feynman diagram that has a single vertex
with $N$ external legs and no propagators.  The
contribution of this cell to $\cL a_1\cdots a_N\cR_e$ is given by 
$\int_C I$ where $I$ is the same integrand
that appears in the expression for the full off-shell amplitude.

Now suppose we have a cell $C_1$ corresponding to a Feynman diagram 
in which we have a single propagator. This could either connect 
two legs of a single vertex 
or two vertices. 
Riemann surfaces in this cell have a natural relation to the
Riemann surfaces corresponding to the vertex (pair of vertices) 
that remains when the propagator is removed. The former is obtained by sewing 
two parts of  
the latter
Riemann surface (pair of Riemann surfaces)  using the plumbing fixture relation
\be 
z\, w = e^{-s+i\theta}, \qquad 0\le s<\infty, \quad 0\le \theta<2\pi\, .
\ee
Here $z$ and $w$ are the local holomorphic coordinates around the extra punctures
that arise due to the removal of the propagator. The moduli of the original Riemann
surface labelling points inside the cell $C_1$ 
can be labelled by the moduli of the cell $\wt C_1$ 
corresponding to the vertex (pair of vertices), and the variables
$(s,\theta)$. 

Now take the original Feynman diagram and replace the propagator
by the propagator of the light states. This contribution will be expressed in terms 
of the
product of the propagator of the light states given by
$(k^2)^{-1} \langle\chi_\alpha^c|c_0^- b_0^+\GG |\chi_\beta^c\rangle$ 
and the contribution to 
the off-shell amplitude from the constituent vertex (pair of vertices). 
The latter can be expressed as an
integral over $\wt C_1$, while the  $(k^2)^{-1}$ term in the propagator can be
expressed as\footnote{The $\alpha'/2$ factor multiplying $k^2$ has been fixed
using the fact that $k^2$ appears in the expression for $L_0^+$ in the combination
$\alpha' k^2/2$.}
\be\label{erep}
{\alpha'\over 4\pi} \int_0^\infty ds \int_0^{2\pi} d\theta \, e^{-s\,\alpha' k^2/2}\, .
\ee
When the propagator connects two vertices, $k$ 
is fixed in terms of the external momenta using momentum conservation. When
the propagator connects two legs of a single vertex, we 
carry out the integration over momenta $k$ treating it as a gaussian integral. In either case,
since the coordinates of $\wt C_1$ together with $(s,\theta)$ give the
coordinates of $C_1$, we see that 
the net contribution has been expressed as an integral over the cell $C_1$.
The integrand of course is
different from the integrand $I$ that appeared in the original off-shell amplitude.
Let us call this integrand $I_1$. Then the contribution to  $\cL a_1\cdots a_N\cR_e$ 
from the cell $C_1$ is given by $\int_{C_1} (I-I_1)$. The subtraction term removes the 
contribution of light states from the propagator  and renders the
integral free from divergences in the $s\to\infty$ limit.

Next consider the case of a Feynman diagram with two
propagators, which we shall label by 1 and 2. 
We define $I$ to be the integrand appearing in the original amplitude, $I_r$ for $r=1,2$
to be the
integrand that appears when we replace the $r$-th propagator by 
the light state propagator and
use \refb{erep}, and $I_{12}$ to be the
integrand that appears when we replace both the propagators by light state propagators 
and use \refb{erep}. 
The contribution to $\cL a_1\cdots a_N\cR_e$ from the cell $C_{12}$ associated with this
Feynman diagram 
is given by $\int_{C_{12}} (I-I_1-I_2+I_{12})$, with the additive factor of $I_{12}$ compensating 
for the over subtraction that we have made by subtracting $I_1$ and $I_2$ from
$I$.

The general procedure is now clear. Contribution to $\cL a_1\cdots a_N\cR_e$ from a
cell $C_{12\cdots k}$ associated with a Feynman diagram with $k$ propagators labelled by
$1,\cdots k$ is given by
\be 
\int_{C_{12\cdots k}} \, \sum_{\ell=0}^k (-1)^\ell \sum_{\{r_1,\cdots r_\ell\}\subseteq \{1,2,\cdots k\}}
I_{{r_1}\cdots r_{\ell}}\, ,
\ee
where $I_{r_1\cdots r_\ell}$ is the integrand that we get by replacing the 
propagators $r_1,\cdots r_\ell$ by light state propagators and then applying 
\refb{erep}.

The procedure described above gives an expression for $\cL a_1\cdots a_N\cR_e$ that is
free from all divergences as long as the energies of external states $a_1,\cdots a_N$ are
below the threshold of production of heavy states. Above this threshold we shall get
divergences from the region where one or more Schwinger parameters $s$ appearing in
\refb{erep} become large 
(see {\it e.g.} \cite{1607.06500}). 
As described in
\S\ref{s6}, this may be
avoided by working with a different effective action in which only states above the threshold
are integrated out. In this case we carry out the algorithm described above by 
including in the subtraction terms not just the 
light state propagators but the projection of the full propagator to states up to a fixed 
mass level.

\sectiono{Removing spurious fields} \label{s5}

Since the effective action has only the light fields as its degrees of 
freedom, one would expect the number of fields to be finite. This is indeed true in the
NS sector since there are no bosonic zero modes which can act on a light state to
create infinite number of light states. However in the R sector we have zero modes of
$\gamma$ in the picture number $-1/2$ sector and zero modes of $\beta$ in the
picture number $-3/2$ sector. Therefore 
we can create infinite number of states at mass level
zero by acting with these oscillators. We shall now show that only a 
finite number of these states
appear in the computation of Feynman diagrams.

Let us focus on the right-moving Ramond sector state, the
analysis for the left-moving Ramond sector
will be identical. 
We drop all reference to the
left-moving sector and also the momentum labels and the space-time spinor index
coming from the right-moving sector. 
The only
exception will be the zero modes of the $b$, $c$, $\bar b$, $\bar c$ 
ghosts since for these the
condition \refb{ech} couples the two sectors. Denoting by $|p\rangle$ 
the Ramond ground state of picture number $p$ in this convention,
the unwanted modes of $|\Phi\rangle$ are the coefficients of
$(\gamma_0)^n c_1|-1/2\rangle$ and $(\gamma_0)^n c_0^+c_1|-1/2\rangle$ for 
$n\ge 1$.
Their  duals, in the sense
described in \refb{echic}, are $c_0^+ (\beta_0)^n c_1|-3/2\rangle$ and 
$ (\beta_0)^n c_1|-3/2\rangle$ respectively.
Now \refb{efa1mod}, \refb{efa2mod} 
show that the propagator 
involves  $b_0^+ \GG$ acting on the
dual basis state.  The $b_0^+$ annihilates the state 
$ (\beta_0)^n c_1|-3/2\rangle$. On the other hand
acting on $c_0^+ (\beta_0)^n c_1|-3/2\rangle$,  $b_0^+\GG$ will
generate $\GG (\beta_0)^n c_1|-3/2\rangle$. 
This is a state carrying right-moving picture number $-1/2$ and ghost number
$(1-n)$.
It is easy to see that this vanishes
for $n\ge 1$ -- there is simply no candidate state with the right ghost number and
picture number at
mass level zero. 
This shows that the fields associated with
states of the form described above do not
appear as intermediate states in the Feynman diagrams. 
Similar arguments together with \refb{ess}, \refb{edde} show that these fields also
do not contribute to $\{S_e,S_e\}$ or $\Delta S_e$.
Therefore for all
practical computation we can drop these fields and their dual, and  
work with a finite number of fields.



\end{document}